\documentclass[12pt]{iopart}
\usepackage{graphicx}

\begin{document}

\title{The effect of shear and bulk viscosities on elliptic flow }
\author{G. S. Denicol${}^{a}$, T. Kodama${}^{b}$ and T. Koide${}^{c}$ }

\begin{abstract}
In this work, we examine the effect of shear and bulk viscosities on
elliptic flow by taking a realistic parameterization of the shear and bulk
viscous coefficients, $\eta$ and $\zeta$, and their respective relaxation
times, $\tau_{\pi}$ and $\tau_{\Pi}$. We argue that the behaviors close to
ideal fluid observed at RHIC energies may be related to non-trivial
temperature dependence of these transport coefficients.
\end{abstract}

\address{$^{a}$Institute f\"ur Theoretische Physik, Johann Wolfgang
Goethe-Universit\"at, Max-von-Laue Str. 1, 60438, Frankfurt am Main, Germany}
\address{$^{b}$Instituto de F\'{\i}sica, Universidade Federal do Rio de Janeiro, C.
P. 68528, 21945-970, Rio de Janeiro, Brasil} 
\address{$^{c}$Frankfurt Institute for Advanced Studies, Johann Wolfgang
Goethe-Universit\"at, Ruth-Moufang Str. 1, 60438, Frankfurt am Main, Germany}

\section{Introduction}

Dissipative hydrodynamics has been applied to relativistic heavy ion
collisions to understand the viscous nature of the quark gluon plasma (QGP),
in particular, to explain the experimental data of the collective flow
parameter $v_2$ by including the effect of shear viscosity \cite{Romatschke}.

One of the objectives of these attempts is to determine $\eta /s$, where $%
\eta $ is the shear viscosity coefficient and $s$ is the entropy density, by
comparing hydrodynamical calculations of $v_{2}$ with the measured values.
However, in addition to the uncertainties on the initial condition for
hydrodynamics and the breakdown of the traditional Cooper-Frye freeze-out
procedure in dissipative hydrodynamics \cite{dkkm5, monnai}, there are
several difficulties in these attempts. One should note that these studies
are mainly based on three assumptions; 1) the effects of other irreversible
currents, such as the bulk viscous pressure and heat conductivity, are
smaller than those of the shear stress tensor, 2) $\eta/s$ does not strongly
depend on temperature and 3) the effects of shear viscosity on observables
are mostly characterized by one transport coefficient, $\eta /s$.

In this work, we would like to discuss the validity of these assumptions.
First, $\eta /s$ is in general a function of temperature. While the exact
temperature dependence of this ratio is still unknown, its qualitative
behavior is somehow well understood. As was shown in Ref. \cite{Lacey}, $%
\eta /s$ exhibits a complex temperature dependence showing a minimum near
the QCD phase transition. Thus, it is not clear that such a temperature
dependence can be approximately replaced by a constant value.

Second, we remark that the effects of the shear stress tensor (or any other
irreversible current) cannot be characterized only by $\eta$. This might be
the case for a fluid described by the relativistic Navier-Stokes (RNS)
equation. However, as was shown in Refs. \cite{dkkm3,pu}, hydrodynamical
equations which violate causality, such as the RNS equations, are unstable
and cannot be used for any application. To solve the problem of acausality
and instability, we have to introduce (at least) one more transport
coefficient called the relaxation time $\tau _{R}$, which represents the
retardation effect in the formation of the irreversible current. Then it is
natural to expect that the relaxation time should affect the behavior of $%
v_{2}$ or any other observable.

Because of the relaxation time, a large $\eta $ does not necessarily induce
a large shear stress tensor. To illustrate this, let us consider the
retardation effect of an arbitrary irreversible current $J$ which is induced
by a thermodynamic force $F$. For the sake of simplicity, we consider a
constant $F$. Then, the current is given by 
\begin{equation}
J=DF(1-e^{-t/\tau _{R}}),
\end{equation}%
where $D$ and $\tau _{R}$ are the transport coefficients for this respective
irreversible current. One can see that if $\tau _{R}$ is small enough
compared to the typical scale of the hydrodynamic evolution, $J$ is simply
proportional to $DF$. However, if $\tau _{R}$ is not small, we have to
consider the competition of $D$ and $\tau _{R}$ to quantify the dissipative
effects.

As a matter of fact, the time scale of the hydrodynamic evolution in
relativistic heavy-ion collisions is of the order of 10 fm. On the other
hand, as is shown in Ref. \cite{knk}, the relaxation time of the shear
viscosity $\tau _{\pi }$ can also reach values around 10 fm. In this sense,
we should not ignore the effect of $\tau _{\pi }$ in the analysis of $v_{2}$.

In this paper, we calculate the effect of the shear and bulk viscosities on
elliptic flow by taking a realistic parameterization of the shear and bulk
viscous coefficients, $\eta$ and $\zeta$, and the corresponding relaxation
times, $\tau_{\pi}$ and $\tau_{\Pi}$. We argue that the behaviors close to
ideal fluid observed at RHIC energies may be related to non-trivial
temperature dependence of these transport coefficients.

\section{Model}

In this work, we will use the memory function method \cite{dkkm,dkkm4},
which is one of the formulations of relativistic dissipative hydrodynamics.
For a general metric, the conservation of energy and momentum is expressed
as 
\begin{equation}
D_{\mu }T^{\mu \nu }=0, 
\end{equation}
where $D_{\mu }$ denotes the covariant derivative. We use the Landau
definition of the four-fluid velocity $u^{\mu }$ in which case the
energy-momentum tensor is decomposed as $T^{\mu \nu }=\varepsilon u^{\mu
}u^{\nu }-\left( p+\Pi \right) \Delta ^{\mu \nu }+\pi ^{\mu \nu }$, where, $%
\varepsilon $, $p$, $\Pi $ and $\pi ^{\mu \nu }$ are the energy density,
pressure, bulk viscous pressure and shear stress tensor, respectively. The
projection operator is defined as usual, $\Delta ^{\mu \nu }=g^{\mu \nu
}-u^{\mu }u^{\nu }$.

In the memory function method, any irreversible current $J$ is induced by a
thermodynamic force $F$ as 
\begin{equation}
\tau _{R}u^{\mu }D_{\mu }J+(1+\tau _{R}D_{\mu }u^{\mu })J=F.
\end{equation}%
This equation is derived to take into account time retardation effects and
the extensivity of hydrodynamic variables. One can regard this equation as
the generalized version of the Maxwell-Cattaneo equation. It should be
emphasized that the existence of the non linear term $\tau _{R}D_{\mu
}u^{\mu }$ is essential to obtain a stable theory as is discussed in Ref. 
\cite{dkkm4}.

By applying this equation to the cases of the shear stress tensor and the
bulk viscous pressure, we obtain 
\begin{eqnarray}
\tau _{\pi }\Delta ^{\mu \nu \lambda \rho }u^{\alpha }D_{\alpha }\pi
_{\lambda \rho }+\pi ^{\mu \nu } &=&\eta \sigma ^{\mu \nu }-\tau _{\pi }\pi
^{\mu \nu }\theta , \\
\tau _{\Pi }u^{\alpha }D_{\alpha }\Pi +\Pi  &=&-\zeta \theta -\tau _{\Pi
}\Pi \theta ,
\end{eqnarray}%
where $\sigma ^{\mu \nu }=D^{<\mu }u^{\nu >}$ and $\theta =D_{\mu }u^{\mu }$
are the thermodynamic forces. We use the traditional notation $%
A^{\left\langle \mu \nu \right\rangle }=\Delta ^{\mu \nu \alpha \beta
}A_{\alpha \beta }$ where $\Delta ^{\mu \nu \alpha \beta }=\frac{1}{2}\left(
\Delta ^{\mu \alpha }\Delta ^{\nu \beta }+\Delta ^{\mu \beta }\Delta ^{\nu
\alpha }-\frac{2}{3}\Delta ^{\mu \nu }\Delta ^{\alpha \beta }\right) $ is
the symmetric traceless double projection operator.

The above hydrodynamical equation are solved numerically in three spatial
dimensions. We use the Smoothed Particle Hydrodynamics (SPH) method. See
Refs. \cite{SPH-heavy ion,dkkm4} for details. We use approximately 80000 SPH
particles with a smoothing parameter $h=0.6$ fm \cite{SPH-heavy ion}.

For simplicity, we use a factorized initial energy density profile into its
longitudinal and transverse parts, as proposed by Ref. \cite{HiranoIC}. The
transverse part is parameterized as in Ref. \cite{dkkm5}. The longitudinal
part is taken as in Ref. \cite{HiranoIC}, with the choice of parameters $%
\eta _{Gaus}=0.8$ and $\eta _{flat}=4$. The initial value of the
irreversible currents will be taken as zero and the impact parameter is
always set as $b=7.5$ fm. We use the equation of state (EoS) introduced in 
\cite{dkkm5}. \ As for the freeze out, we used the ordinary Cooper-Frye
formula \cite{SPH-heavy ion}. However, we did not consider non-equilibrium
corrections to the one-particle distribution function, because, as was
pointed out in Refs. \cite{dkkm5,monnai}, this correction gives rise to
unphysical behaviors and is not reliable. However, this will not affect the
conclusions of this paper.

\begin{figure}[tbp]
\begin{minipage}{.5\linewidth}
\includegraphics[scale=0.3]{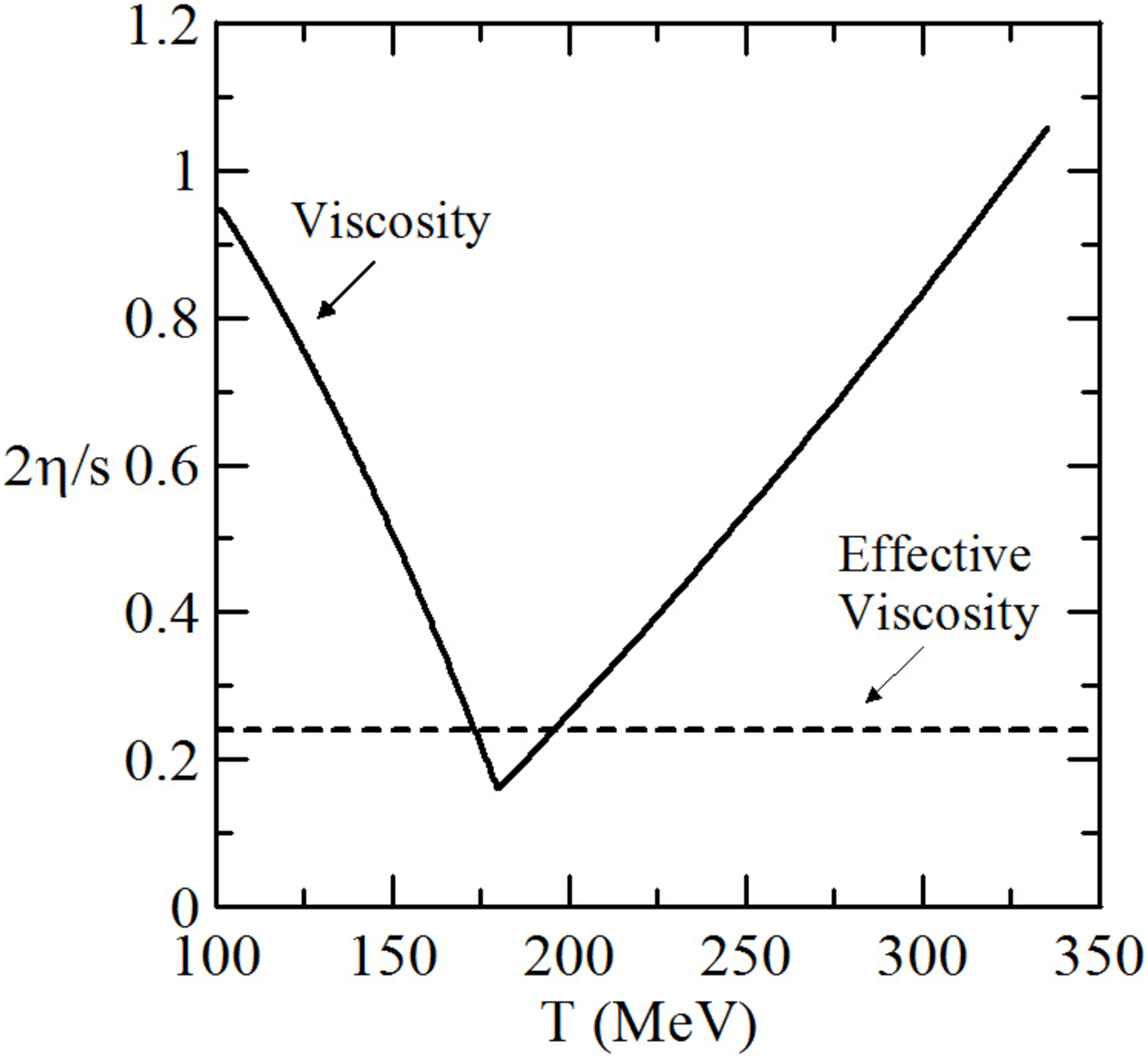}
\label{Shear_Coef}
\end{minipage}
\begin{minipage}{.5\linewidth}
\includegraphics[scale=0.3]{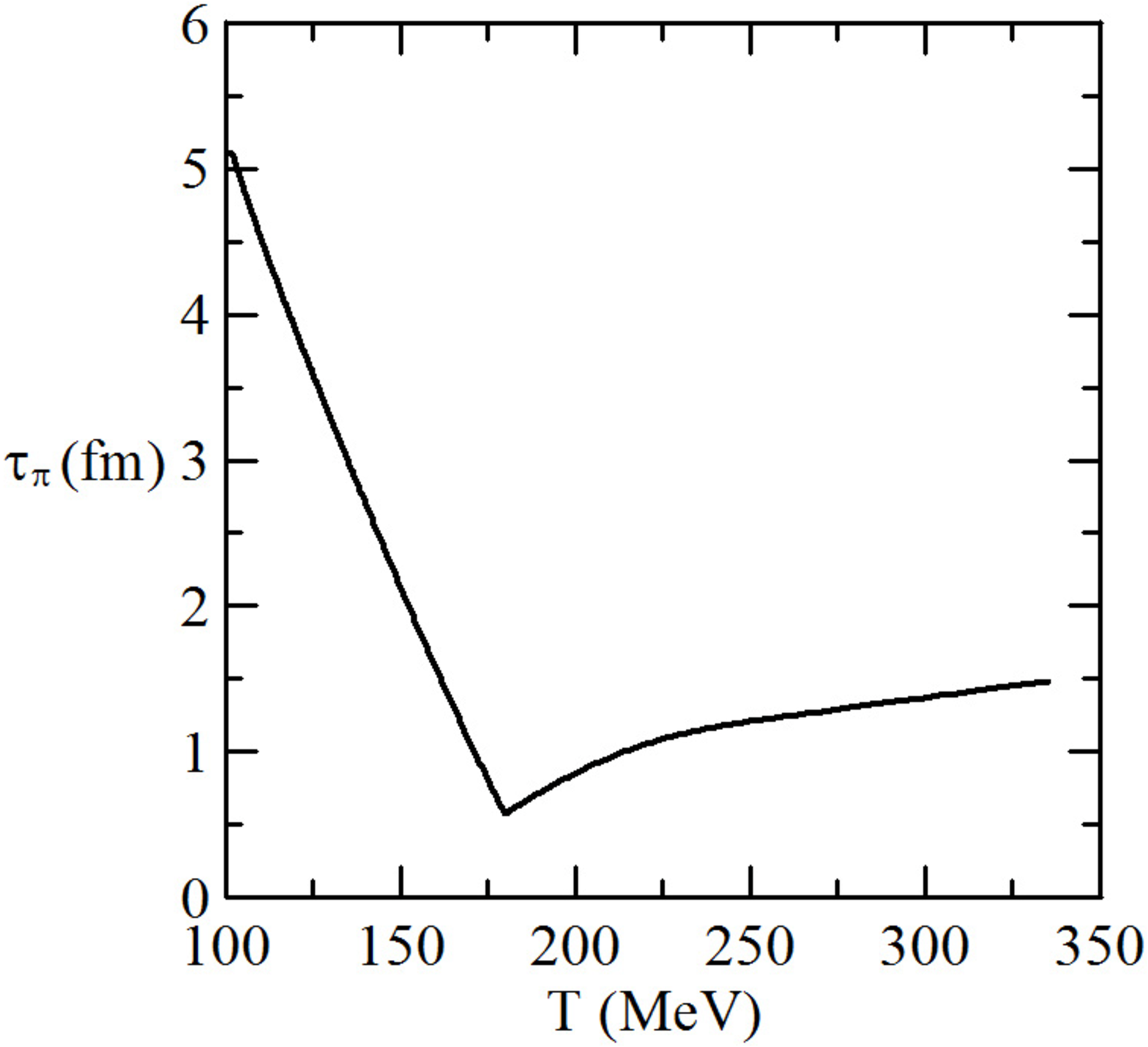}
\label{Shear_Tau}
\end{minipage}
\caption{Our parameterization of the shear viscosity $\protect\eta/s$(left
panel) and the shear relaxation time $\protect\tau_{\protect\pi}$ (right
panel).}
\end{figure}

As mentioned in the introduction, the temperature dependence of the
transport coefficients is very important. The microscopic formula to
calculate these quantities is given in Ref. \cite{knk}. However, the
application of this formula is still limited to leading order perturbative
calculations. Fortunately, in this approximation $\eta $ and $\zeta $ become
equivalent to the ones calculated in the Green-Kubo-Nakano formalism and we
can use these results. In this paper, we use the lattice QCD calculations 
\cite{Nakamura} for the QGP phase and we use the results from Ref. \cite%
{Noronha} for the hadron phase. These calculations for $\eta /s$ were fitted
and the result is shown in Fig. \ref{Shear_Coef}(a). We can see that $\eta/s$
shows a minimum around $T\sim 180$ MeV. For the bulk viscosity coefficient $%
\zeta /s$, we use the parameterization showed in Ref. \cite{dkkm5} with a
critical temperature of $T\sim 180$ MeV.

Similarly, in leading order calculations, the microscopic formula predicts a
simple relation for the shear viscosity, $\tau _{\pi }=\eta/P$. For the bulk
viscosity, there is still no result and we simply assume $\tau _{\Pi
}=9\zeta /(\epsilon -3P)$. Since $\tau _{\pi }$ ($\tau _{\Pi }$) is linear
with the transport coefficient $\eta $ ($\zeta $), the critical behavior of $%
\eta $ ($\zeta $) is transfered to that of $\tau _{\pi }$ ($\tau _{\Pi }$).
Thus, $\tau _{\pi }$ will have a minimum around the QCD phase transition
while $\tau _{\Pi }$ will display a maximum. In this sense, the complex
temperature dependences of the relaxation times are induced by those of $\eta
$ and $\zeta$.

\section{Results}

\begin{figure}[tbp]
\begin{minipage}{.5\linewidth}
\includegraphics[scale=0.3]{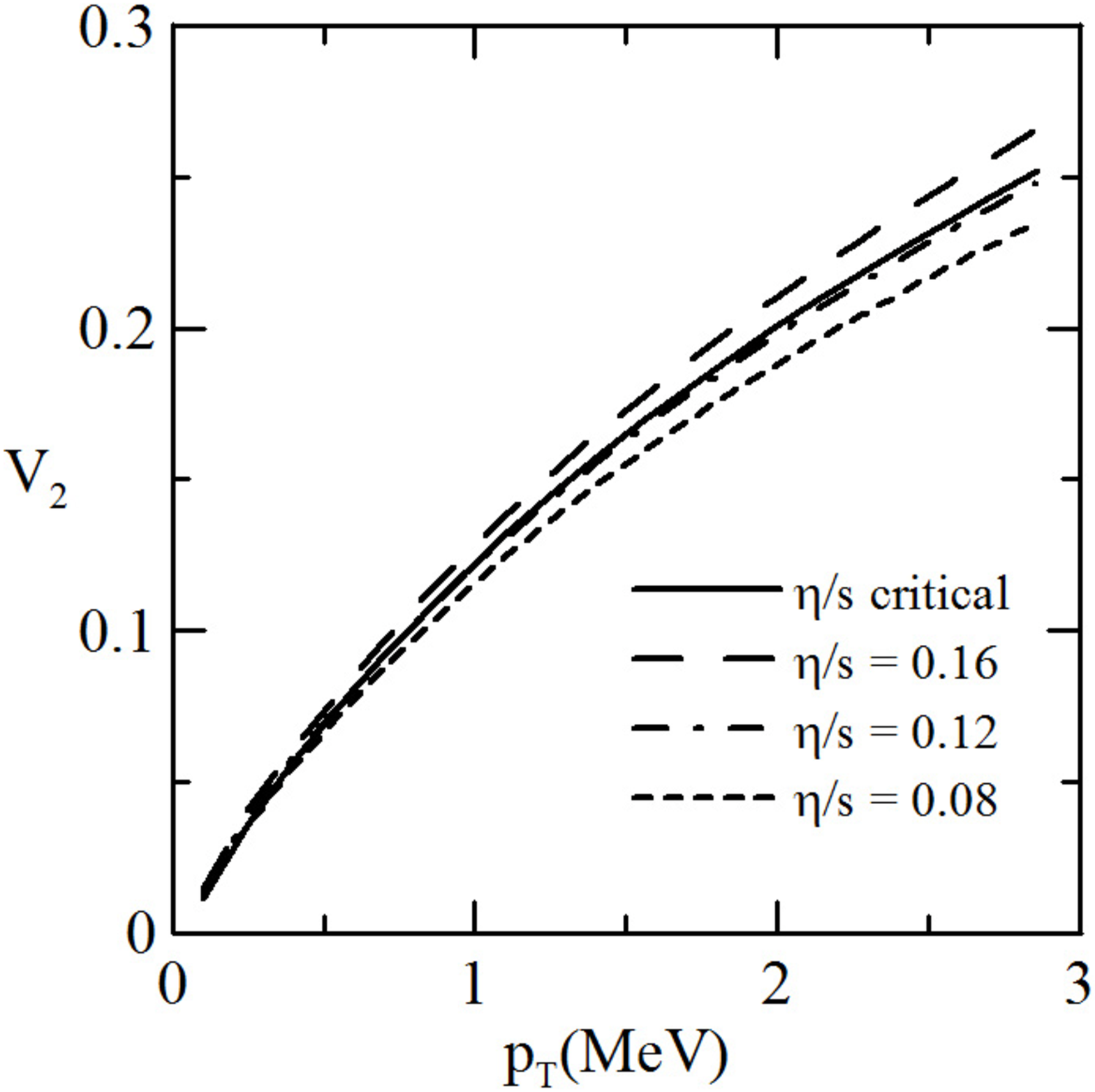}
\label{Effective_Vis}
\end{minipage}
\begin{minipage}{.5\linewidth}
\includegraphics[scale=0.3]{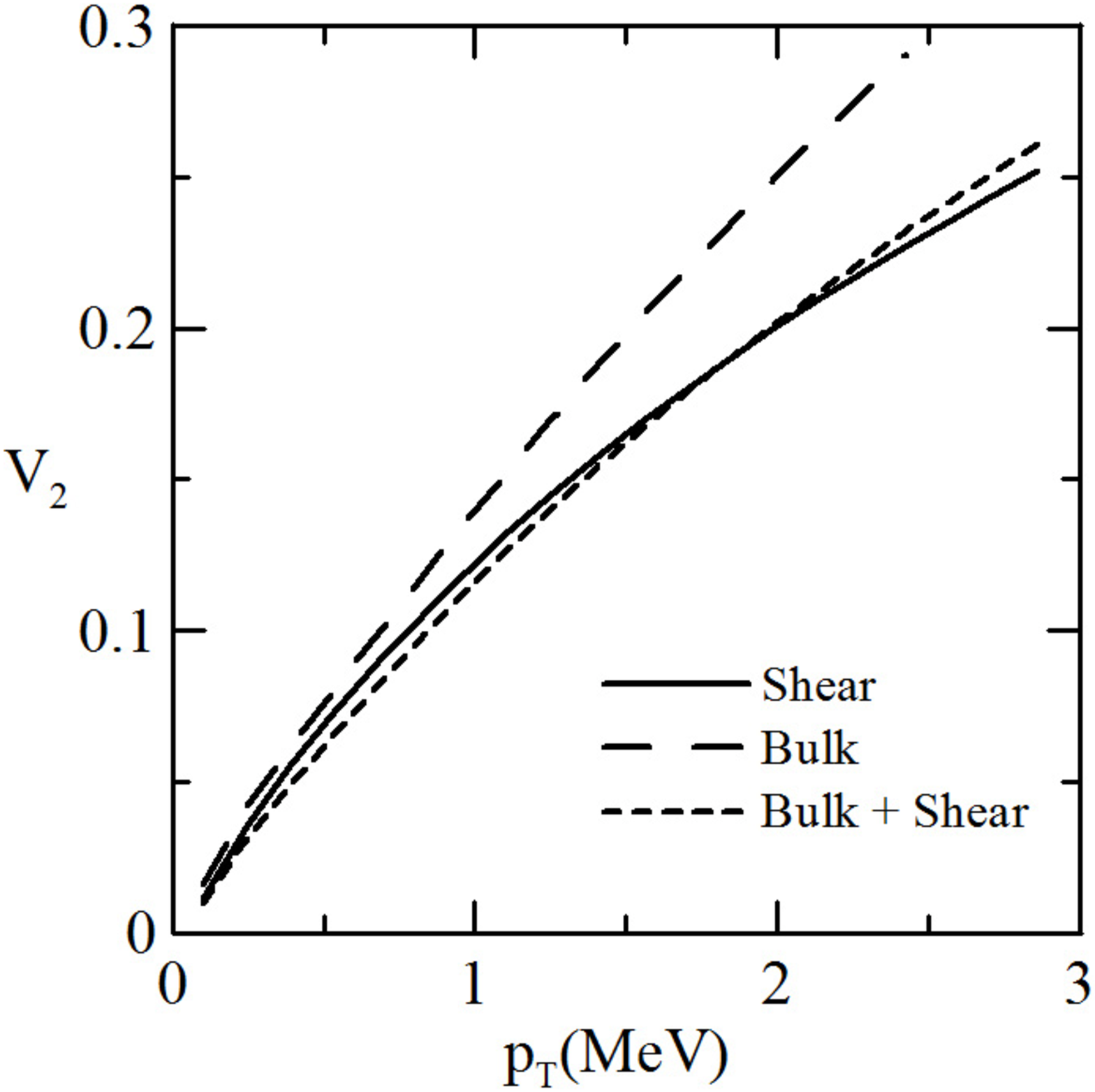}
\label{bulk}
\end{minipage}
\caption{The elliptic flow $v_{2}(p_{T})$ calculated for different choices
of $\protect\eta/s$ (left panel) and calculated including the effects of
bulk viscosity (right panel). }
\end{figure}

In Fig. \ref{Effective_Vis}(a), we show $v_{2}(p_{T})$ for different choices
of $\eta/s$ and neglecting bulk viscosity. These calculations were made for
thermal pions with a freeze-out temperature of $T_{f}=130$ MeV. The solid
line corresponds to the case where $\eta /s$ has the temperature dependence
shown in Fig. \ref{Shear_Coef}(a) whereas the dashed, dashed-dotted and
dotted lines correspond to constant values of $\eta /s$ of 0.08, 0.12 and
0.16, respectively. One can see that the elliptic flow calculated with the
temperature dependent $\eta /s$ can be well reproduced by the one calculated
with an effective constant shear viscosity coefficient $\eta_{eff} /s=0.12$.

The quantity $\eta_{eff}/s$ is very close to the minimum of $\eta/s$. This
can be explained by the behavior of the relaxation time, shown in Fig. \ref%
{Shear_Tau}(b). As mentioned in the introduction, $\tau _{\pi }$
characterizes the time scale for the velocity gradients to be converted to
viscosity. Because the fluid expands with comparable time scale with $%
\tau_{\pi}$, the magnitude of the shear stress tensor does not become large
despite the large values of $\eta/s$ achieved throughout the hydrodynamical
evolution.

It should be noted that $\eta_{eff}/s$ is not bounded by the minimum of $%
\eta/s$. Let us consider another set of parameters, $\tau _{\pi }=3\eta /P$.
In this case, the effective viscosity $\eta _{eff}/s$ becomes smaller than
the minimum value of $\eta /s$. This result shows that the value of $%
\eta_{eff}/s$ is not a direct measurement of $\eta /s$. It can come from the
combined effect of $\eta /s$ and $\tau_{\pi}$, and the small values of $%
\eta_{eff}/s$ found at RHIC can just as well be a manifestation of the large
relaxation time in the hadronic and QGP phases. Whether $\eta_{eff}/s$ can
be used to quantify the value of viscosity at the phase transition or not,
depends on the magnitude of the relaxation time in these regions.

In Fig. \ref{bulk}(b), we show $v_{2}(p_{T})$ considering the combined
effects of bulk and shear viscosities, with realistic transport
coefficients. The solid line corresponds to the calculation with only shear
viscosity, the dashed line to the calculation with only bulk viscosity and
the dashed-dotted line to the calculation with both bulk and shear
viscosities. We see that the effect of bulk viscosity is rather small
despite the large values of $\zeta/s$ near the phase transition. This can be
explained as an effect of the relaxation time $\tau _{\Pi }$, again. Since $%
\tau _{\Pi }$ also becomes large there, the values of $\zeta /s$ do not
generate a large bulk viscous pressure. Since $\zeta $ decreases very
quickly outside the phase transition region, we cannot see considerable
effects of the bulk viscosity.

We remark that all the calculations showed in this work were done in three
spatial dimensions. However, we were mainly interested in $v_{2}$ at zero
rapidity, where we confirmed that the Bjorken scaling $\mathit{ansatz}$ is a
good approximation. Thus, all the results showed here are approximately
equivalent to (2+1)D hydrodynamical calculations with the Bjorken scaling
for the longitudinal direction. This happened mainly because of our choice
of initial condition, which displays a plauteu in energy density in the
longitudinal direction. Studies with different initial conditions, specially
in the longitudinal direction, will be done in future.

\section{Concluding remarks}

We have shown that due to the interplay between the effects of viscosity and
relaxation time, it is possible to fit the behavior of collective flow in
terms of one unique effective shear viscosity $\eta _{eff}/s$. However, its
value depends critically on the evolution of the system. Specifically, the
value is affected by the time span for which the system remains in different
phases. For the LHC\ energies, for example, we would expect a different
value for the effective viscosity since the system will stay longer in the
quak gluon plasma phase. However, the details of such changes are hard to
calculate due to the uncertainties on $\eta $. We also remark that the
uncertainties on the relaxation time are even larger since not so many
effort have been applied to calculate the temperature dependence of this
transport coefficient (unlike the $\eta /s$ case).

We acknowledge discussions with J. Noronha, P. Huovinen, R. Lacey and D. H.
Rischke. This work has been supported by CNPq, FAPERJ, CAPES, PRONEX and the
Helmholtz International Center for FAIR within the framework of the LOEWE
program (Landesoffensive zur Entwicklung Wissenschaftlich- Okonomischer
Exzellenz) launched by the State of Hesse.

\end{document}